# Observation of O+ Characteristics During the Terrestrial Alfvén Wing State Induced by the April 2023 Coronal Mass Ejection

Haoming Liang[1,2], Li-Jen Chen[2], Stephen A. Fuselier[3,4], Roman G. Gomez[3], Brandon Burkholder[2,5], Naoki Bessho[1,2], Harsha Gurram[1,2], Rachel C. Rice[1,2], Jason Shuster[6], Akhtar S. Ardakani[6]

[1]Department of Astronomy, University of Maryland, College Park, MD, USA
[2]NASA Goddard Space Flight Center, Greenbelt, MD, USA
[3]Southwest Research Institute, San Antonio, TX, USA
[4]University of Texas at San Antonio, San Antonio, TX, USA
[5]Goddard Planetary Heliophysics Institute, University of Maryland Baltimore County, Baltimore, MD, USA
[6]Space Science Center, University of New Hampshire, Durham, NH, USA

**Abstract**

We report Magnetospheric Multiscale observations of oxygen ions (O+) during a coronal mass ejection in April 2023 when the solar wind was sub-Alfvénic and Alfvén wings formed. For the first time, O+ characteristics are studied at the contact region between the unshocked solar wind and the magnetosphere. The O+ ions show energies between 100s eV and ~30 keV. The possible sources are the ring current, the warm plasma cloak, and the ionosphere. The O+ ions exhibit bi-directional streaming along newly-formed closed field lines (CFLs), and dominantly anti-parallel on earlier-formed CFLs. Escaping O+ ions in the unshocked solar wind are observed. During the recovery phase, the O+ pitch-angle distribution associated with flux tubes shows dispersion, indicating potential loss to the solar wind. Our results show escaping as well as trapped O+ ions in the region where a magnetic cloud, an Alfvén wing, and magnetospheric field lines are mixed.

**Plain Language Summary**

During a coronal mass ejection in April 2023, NASA's Magnetospheric Multiscale mission observed oxygen ions (O+) under a rare sub-Alfvénic solar wind condition, which turned Earth's magnetic field into special magnetic structures called Alfvén wings. This study is the first to look at O+ ions in the area where the unshocked solar wind meets Earth's magnetosphere during Alfvén wing state. The O+ ions had energies from a few hundred electron volts (eV) to about 30,000 eV. These O+ ions likely came from different regions around Earth, including the ring current, the warm plasma cloak, and the ionosphere. The O+ ions moved in both directions along newly-formed closed magnetic field lines and mainly in one direction along field lines that formed earlier. Some O+ ions were observed in the solar wind, showing a dispersion pattern that suggests they might be lost to space during the recovery phase of the Alfvén wing state. Overall, the study found both escaping and trapped O+ ions in the area where the field lines of the solar wind, the Alfvén wings, and Earth's magnetosphere mix. This helps us understand how the O+ ions behave during Alfvén wing states and their potential loss to space.

**Key Points**
    (1) O+ ions were bidirectional on newly-formed CFLs and dominantly antiparallel on the CFLs that were formed earlier.
    (2) O+ ions along with energetic protons and electrons escaped from the magnetosphere into the unshocked solar wind.
    (3) O+ pitch-angle dispersion was observed, indicating potential loss to the solar wind during the Alfvén wing recovery.

## 1. Introduction

Understanding the dynamic interaction between the solar wind and Earth's magnetosphere is a complex yet crucial endeavor in space plasma physics. Under normal solar wind conditions, Earth's magnetosphere contains various boundary regions, including the bow shock, the magnetosheath, and the magnetopause. These boundaries are intricately shaped by solar wind parameters such as a Mach number, dynamic pressure, and an interplanetary magnetic field (IMF) [e.g., Kivelson and Russell, 1995]. Typically, the solar wind is super-Alfvénic, with an Alfvén Mach number ($M_A$) of 8 or above. However, in a magnetic cloud (MC) embedded in a coronal mass ejection (CME), the Alfvén Mach number can drop to values of 2 or less [e.g., Lavraud and Borovsky, 2008]. In rare cases, the Alfvén Mach number could drop below 1, which induces a unique phenomenon known as Alfvén wings [e.g., Ridley, 2007]. The Alfvén wings consist of open field lines with one end connected to the Earth.

The unusual conditions within a CME, such as the presence of a low-density magnetic cloud, which leads to sub-Alfvénic flow, cause intricate alterations in the dynamics of Earth's magnetosphere. Chané et al [2012] discuss Geotail observations of Alfvén wings formed during sub-Alfvénic and sub-fast solar wind conditions upstream of the Earth on May 24-25, 2002. Under these conditions, the magnetosphere was unusually quiet, with minimal auroral activity. Chané et al [2015] used 3D MHD simulations to study the same event. They found that during that event, the Earth's closed magnetic field line region became symmetric, extending sunward and shrinking tailward. In their case, the upstream IMF $B_y$ was positive. The configuration of open field lines underwent a transformation, with field lines originating from the northern hemisphere converging towards the dawn Alfvén wing, and those from the southern hemisphere aligning with the opposite wing. According to Chané et al [2015], as the Alfvén wings formed, the tail lobes vanished entirely, and there was a significant reduction in auroral activity, leading to a state of geomagnetic quietness in the magnetosphere. The inner magnetosphere response during the Alfvén wing state was also studied. By using simultaneous multiple spacecraft observations, Lugaz et al. [2016] studied a CME event on January 17, 2013, and revealed a rapid and intense depletion of electrons in the outer radiation belt.

On April 24, 2023, a CME hit Earth's magnetosphere. The interaction between the solar wind and the magnetosphere was observed by MMS in the dayside pre-noon sector near the southern cusp at [10.7, -7.9, -6.8] $R_E$ in GSE coordinates. During this CME, the embedded MC had significantly low density which resulted in the solar wind $M_A<1$ (by increasing the Alfvén speed). When the MC reached Earth's magnetosphere, it led to a period (12:30-14:40 UT) of an interaction between the unshocked sub-Alfvénic solar wind and the magnetosphere. Alfvén wings formed during this period based on observations and simulations [Chen, L.-J. et al., 2024; Chen, Y., et al., 2024; Burkholder et al., 2024].

Prior research indicates that during a typical storm period instigated by CMEs, O+ ions from the ionosphere are accelerated in the cusp, convected over the polar cap into the tail lobes, and then enter the plasma sheet where they undergo further acceleration and transport processes [e.g., Liao et al., 2015]. They are energized to levels ranging from a few tens to hundreds of kilo-electron volts (keV) [e.g., Fok et al., 2006]. These O+ ions significantly contribute to the ring current, leading to plasma pressure enhancements in the inner magnetosphere [e.g., Kronberg et al., 2014].

Despite the forementioned studies, the behavior of the magnetospheric O+ ions during the Alfvén wing state remains unclear. In this study, we investigate O+ characteristics at the contact region between the unshocked solar wind and the magnetosphere for the first time. We use Magnetospheric Multiscale (MMS) observations [Burch et al., 2016] in the magnetic cloud, an Alfvén wing, and the freshly-formed closed field line region during the CME event on April 24, 2023. This study sheds light on the impact of Alfvén wings on O+ circulation in Earth's magnetosphere. It also provides insight into the heavy ion circulation during the Alfvén wing state for extrasolar planets [e.g, Saur et al., 2013] and planetary satellites [e.g., Neubauer,1998; Kivelson et al., 2004].

## 2. Methodology

We use magnetic field data from the Fluxgate Magnetometer (FGM) [Russell et al., 2016], electron and ion data from the Fast Plasma Investigation (FPI) [Pollock et al., 2016], energetic electron data from the Fly's Eye Energetic Particle Spectrometer (FEEPS) [Blake et al., 2016], H+ and O+ data from Hot Plasma Composition Analyzer (HPCA) [Young et al., 2016]. In this event, the bleedover of intense proton (H+) fluxes leads to contaminated O+ flux [Fuselier et al., 2021]. Based on the time-of-flight data for the period 14:05-14:25 (not shown), the O+ flux below 10-keV was significantly affected by the H+ contamination. While it is possible that genuine O+ signals exist below this energy level, we still use the 10-keV as the conservative threshold to ensure that all analyzed O+ signals are uncontaminated and reliable. The O+ contamination below 10 keV from 12:30-14:24 UT and from 14:38-16:18 UT, and below 300 eV from 16:18-16:23 UT has been removed before plotting the moments and the spectra.

## 3. Results
### 3.1 Overview

The primary focus of the paper is to report oxygen ions (O+) characteristics in the region with a mixture of a magnetic cloud, an Alfvén wing, and magnetospheric field lines (14:00-14:40 UT) and in the region with isolated flux tubes in the shocked solar wind (15:00-15:22 UT) right after the Alfvén wing phase.

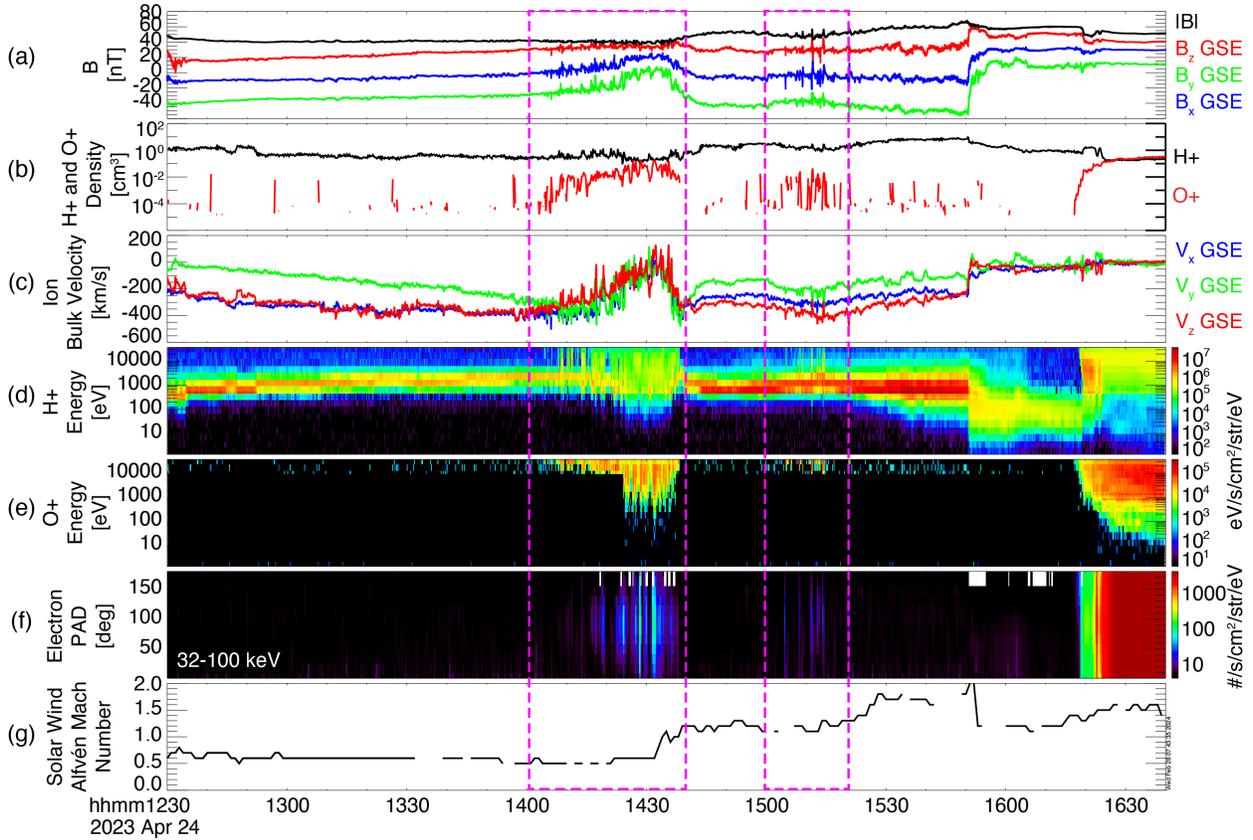

Figure 1 Overview O+ observations in the 2023 April CME event: (a) magnetic field, (b) H+ and O+ densities, (c) ion bulk velocity, (d) H+ energy flux, (e) O+ energy flux, (f) electron PAD for 32-100keV, and (g) the Alfvén Mach number $M_A$ from the OMNI dataset using Wind as the upstream spacecraft propagated to the nose of the bow shock. $M_A$ is the ratio of the solar wind speed and the Alfvén speed. Highlighted intervals are discussed in detail in this paper.

Figure 1 shows an overview of this event, including O+ observations. After 12:30 UT, MMS observed only the unshocked solar wind until 14:05 UT. According to Chen, L.-J., et al. [2024], from 14:05 to 14:38 UT, MMS observed Alfvén wing filaments. The Alfvén wing filaments are characterized by open field lines with one end connecting to the Earth's ionosphere and the other connecting to the solar corona. They were identified by the presence of energized electrons and ions that show no significant enhancement in the upstream solar wind, suggesting that the particles were probably originated from the Earth or generated by magnetic reconnection. A region with primary closed field lines (CFLs) was also captured between 14:25 and 14:35 UT. According to Chen, L.-J., et al. [2024], this region was found to be freshly formed because, different from the magnetospheric steady CFL region at 16:30 UT, the ion flux above 3 keV had a weak intensity, while the increase in electron flux was confined to energies below 5 keV. Furthermore, Gurram et al. [2024] suggest that this CFL region is likely a result of dual-wing reconnection. Despite these differences, the magnetic field components and ion densities were similar to those at 16:30 UT.

Between 14:05 and 14:38 UT, O+ density and energy were enhanced (Figures 1b and 1e). Since the sub-Alfvénic solar wind ends locally at MMS at ~14:40 UT [Chen, L.-J., et al., 2024], this enhancement in O+ was observed during the Alfvén wing state. This increase in O+ ion flux took place on both open and closed field lines.

After ~14:40 UT, the solar wind became super-Alfvénic with low Mach number ($M_A$=1.1-2.0). The dayside magnetopause thus moved away from MMS. In 15:00-15:22 UT, MMS observed a few isolated flux tubes with O+ enhancements during this recovery phase from the Alfvén wing state.

In the following sections, we analyze the characteristics of the O+ in the intervals as highlighted by the dashed boxes in Figure 1. The interval between 14:00-14:40 UT shows O+ characteristics in the region with mixed field lines of the magnetic cloud, the Alfvén wing, and the magnetosphere during the Alfvén wing state. The interval between 15:00-15:22 UT shows the O+ behavior in multiple isolated flux tubes during the recovery phase from the Alfvén wing state.

**3.2 Interval 14:00-14:40 UT**

Figures 2a-l show the observation from 14:00-14:40 UT during the interaction between the unshocked solar wind and the magnetosphere. The O+ density increases as the magnitude of the H+ perpendicular bulk speed gradually decreases from ~400 km/s at 14:00 UT to zero at around 14:25 UT. The 20eV-20keV electron energy fluxes parallel and anti-parallel (Panels e and f respectively) to the local magnetic field are used to distinguish field line connectivity [e.g., Fuselier et al., 2022, 1997; Burkholder et al., 2022]. The detailed analysis in Chen, L.-J., et al., [2024] shows that the magnetic structures including Alfvén wing filaments (i.e., flux tubes with one end connected to the Earth), newly-formed CFLs due to dual-wing reconnection, and MC field lines, are intermittently observed in 14:05-14:24 and 14:37-14:38 UT, while a primarily magnetospheric CFL region is observed from 14:25 to 14:35 UT. The Alfvén wing filaments are generated by the reconnection between the MC field lines and the CFLs occurring either north or south of the MMS. The observation of O+ ions commonly indicates the connection of the magnetic field line to the Earth; however, it is important to note that on the spatiotemporal scales of transient structures smaller than the O+ gyro-motion, O+ ions may also be observed on the MC field lines. Between 14:05-14:24 UT, the O+ flow was separated from the H+ flow. One possible explanation is that they were from different sources and distinct physical processes they experienced while escaping from the magnetosphere [e.g., Möbius et al., 1986; Fuselier et al., 1995; Fuselier, 2020]. Specifically, the observed H+ included both the solar wind and the magnetosphere origins, while the O+ ions were from the magnetosphere. Given the significant speed of O+ perpendicular flow (ranging between 0~200 km/s), we convert the O+ energy spectra and pitch angle distribution (PAD) to the frame where $V_\perp(O+) = 0$ (similar to the $V_\perp(H+) = 0$ frame in Fuselier et al [2019a]), as shown in Panels g and h for O+ omni energy flux and PAD, respectively. The O+ energy in the $V_\perp(O+) = 0$ frame ranges from 1-30 keV (Panel g), consistent with the combined warm plasma cloak (up to a few keV) [Chappell et al., 2008] and ring current populations. The observed O+ population could also include energized O+ by reconnection at the north or south of the spacecraft along the field lines.

Figure 2 Observations between 14:00-14:40 UT (a)-(l) and zoom-in view between 14:23-14:26 UT (m)-(u): (a) magnetic field, (b) H+ and O+ densities, (c) H+ and O+ bulk velocities perpendicular to the local magnetic field, (d) ion energy spectrum, (e) parallel and (f) anti-parallel electron energy flux to the local magnetic field, (g) O+ omni energy flux (h) O+ PAD (i) O+ parallel (pitch-angle = 0°–60°), (j) anti-parallel (pitch-angle = 120°–180°), and (k) perpendicular (pitch-angle = 60°–120°) energy flux to the local magnetic field, (l) the anti-parallel energy flux subtracted by the parallel energy flux. (g)-(l) are in the $V_\perp(O+) = 0$ frame. Black dashed boxes highlight the Interval A and B for zoom-in view. The Panels (m)-(u) are the zoom-in view of Panels (d)-(l) in Interval A (14:23-14:26 UT). The red (blue) dashed boxes highlight the interval with newly-formed (long-formed) closed field lines.

Although the O+ PAD shows bidirectional field-aligned distributions, the fluxes between parallel and anti-parallel are asymmetric on most of the open and CFLs. The asymmetric bidirectional signature can be due to a combination of different sources or the gradual loss of the populations. Since some of the Alfvén wing filament field lines are newly open due to reconnection [Chen, L.-J., et al., 2024], O+ can be gradually lost along the field lines. The timescale of trapped O+ loss along the filament field lines is proportional to the bouncing period before the CFLs are open. As a rough estimate, the bouncing period between the northern and southern mirror points of a 10-keV O+ ion with a 30-degree pitch angle at the equatorial plane L=10 is about 12 minutes [Jordanova et al., 2020]. Therefore, an O+ ion at the equatorial plane makes its way to one mirror point in 3 minutes (it is 6 minutes if the O+ starts from the cusp region of the opposite hemisphere). This estimate provides the timescale for an O+ ion to be lost on a newly open field line. Since the Alfvén wing filament intervals are much shorter than the O+ escaping time, it is reasonable that MMS still observes the bidirectional O+ PAD (similar to trapped O+ signatures) on the Alfvén wing filaments in Figure 2 between 14:05-14:24 UT.

The O+ energy spectrum and PAD in the interval between 14:25-14:35 UT may indicate multiple O+ sources. In this interval, $V_\perp(O+) = V_\perp(H+) = 0$, and CFLs are dominant [Chen, L.-J., et al., 2024; Gurram et al., 2024]. The lower limit of the O+ energy in the $V_\perp(O+) = 0$ frame drop to ~100 eV, which is significantly lower than that observed in the convecting magnetic structures between 14:05-14:24 UT. The warm plasma cloak and ionospheric O+ outflow may have a contribution to the low energy population (<1 keV). The O+ PAD between 14:25-14:35 UT shows counter-streaming signatures with broader pitch-angle ranges at the parallel and anti-parallel directions than those observed between 14:05-14:24 UT. It is worth noting that the observed lower limit of O+ energy (~100 eV) is higher than that in the regions with the steady magnetospheric close field lines at 16:30 UT (~20 eV) as shown in Figure 1.

We compare O+ flux between 14:05-14:24 UT and 14:25-14:35 UT. Figures 2i, j, and k show the O+ energy flux parallel (pitch-angle = 0°–60°), anti-parallel (pitch-angle = 120°–180°), and perpendicular (pitch-angle = 60°–120°) to the local magnetic field in the $V_\perp(O+) = 0$ frame, respectively. The selection of pitch-angle range is based on the O+ PAD enhancement between 14:05-14:24 UT. Panel l is the anti-parallel energy flux subtracted by the parallel energy flux; thus, the red (blue) color indicates the dominant parallel (anti-parallel) energy flux at various energy channels. Panel l shows a mixture of parallel and anti-parallel flux between 14:05-14:24 UT and a dominant anti-parallel O+ flux between 14:25-14:35 UT. To gain deeper insights, we investigate the O+ characteristics in Intervals A and B.

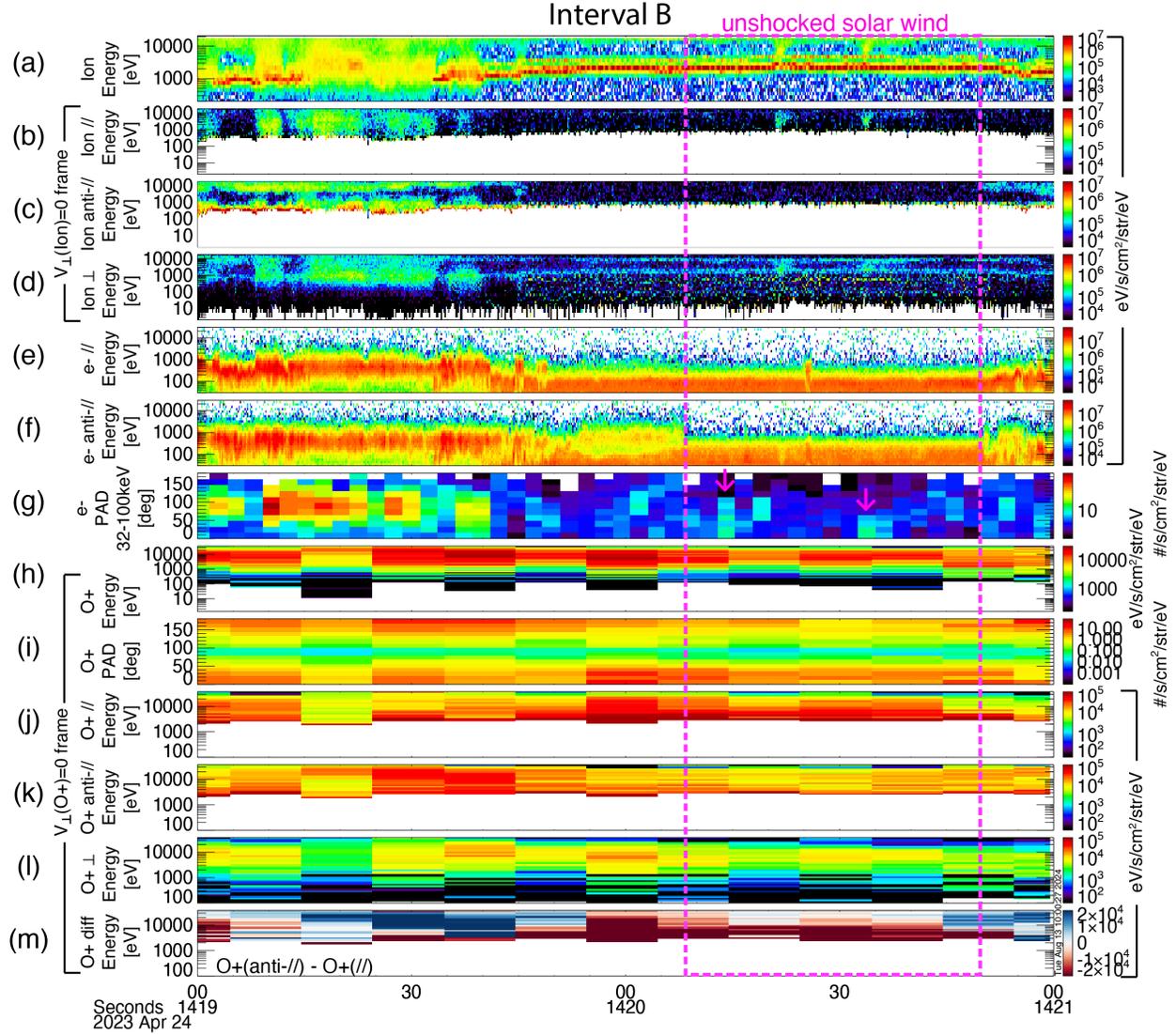

Figure 3 Observations in Interval B (14:19-14:21 UT) from Figure 2: (a) ion energy spectrum, (b) ion parallel (pitch-angle = 0°–45°), (c) anti-parallel (pitch-angle = 135°–180°), and (d) perpendicular (pitch-angle = 67.5°–112.5°) energy flux to the local magnetic field, (e) parallel and (f) anti-parallel electron energy flux to the local magnetic field, (g) electron PAD with energy of 32-100 keV, (h) O+ omni energy flux, (i) O+ PAD, (j) O+ parallel (pitch-angle = 0°–60°), (k) anti-parallel (pitch-angle = 120°–180°), and (l) perpendicular (pitch-angle = 60°–120°) energy flux to the local magnetic field, (m) the anti-parallel energy flux subtracted by the parallel energy flux. (b)-(d) are in the $V_\perp(ion) = 0$ frame. (h)-(m) are in the $V_\perp(O+) = 0$ frame. Magenta dashed boxes highlight the unshocked solar wind interval. Magenta arrows highlight the electron jets.

To examine the O+ characteristics on the CFLs, we focus on Interval A (14:23-14:26 UT) as shown in Figures 2m-u. According to Chen, L.-J., et al. [2024], Interval A marks the entry into the CFL region. Selected intervals are highlighted by red and blue dashed boxes. In these intervals, high-energy counter-streaming electrons (up to ~1 keV) are observed without the MC-like low-energy population (e.g., those between 14:23:00-10 UT). The ion energy spectrum (Panel m) shows varying states of deceleration and heating. These variations depend on the duration of CFL formation. In the blue boxes, ions are fully heated, indicating the CFLs were formed earlier. In the red boxes, a cold ion beam (similar to the MC-like population in 14:23:00-10 UT but decelerated) with a heated background suggests newly-formed CFLs. Panel p shows that the O+ energy flux is less intense on the newly-formed CFLs than on the ones formed earlier. Panel u shows a mixture of parallel and anti-parallel fluxes at various energy channels on the newly-formed CFLs, indicating the O+ sources at the north and south of the MMS. Note that the 10-second cadence of the O+ data is significantly longer than the intervals of transient structures, such as Alfvén wing filaments and newly-formed CFLs [Chen, L.-J., et al., 2024]. This discrepancy could result in temporal aliasing, where parallel and anti-parallel O+ fluxes may occur at different times within intervals shorter than 10 seconds. For the CFLs that were formed earlier, the O+ flux is mainly anti-parallel, indicating the O+ population comes from the north of the MMS. The O+ ions propagated to the observed CFL regions from their source locations along the field lines. The O+ ions with pitch angles near the parallel or anti-parallel directions propagated faster than those with pitch angles near the perpendicular direction. Consequently, we observed mainly parallel and anti-parallel O+ fluxes on the newly-formed CFLs, while the earlier-formed CFLs exhibited relatively more O+ fluxes with pitch angles close to the perpendicular direction.

The escaping O+ in the unshocked solar wind is observed in Interval B (14:19-14:21 UT) in Figure 3, highlighted by the magenta dashed box. Panel a shows the cold ion beam (~1-2 keV) as the MC population, while the energetic ion population (6-20 keV) likely originates from the magnetosphere. The ion spectra (Panels b-d) display dominant perpendicular flux with two short parallel flux enhancements. Panel g reveals two energetic parallel electron jets (32-100 keV) at 14:20:15 UT and 14:20:35 UT. A bidirectional O+ flux with dominant parallel flux (2-20 keV in the $V_\perp(O+) = 0$ frame) is seen in Panels h-m. These escaping O+ signatures, along with energetic protons and electrons, resemble observations in the magnetosheath during super-Alfvénic solar wind by Möbius et al. [1986]. However, the energies of the escaping species in our study are lower than those in Möbius et al. [1986] (≥40 keV for ions, 70-207 keV for electrons). This observation updates our knowledge of the energy of escaping O+. The relatively low energy of escaping O+ may also suggest that O+ leakage occurs more easily under unshocked solar wind conditions. The fate of the escaping O+ remains uncertain; they may either propagate with the MC or return to the magnetosphere due to their large gyro-radius and the near-tangential configuration of the MC field lines to the magnetosphere.

### 3.3 Interval 15:00-15:22 UT

As shown in the left column of Figure 4, multiple isolated flux tubes were observed in the interval between 15:00-15:22 UT (marked by the magenta arrows in Panel h1). As discussed in Section 3.1, during this interval, the solar wind became super-Alfvénic ($M_A$ =1.1-2.0) and the magnetosphere contracted (i.e., the dayside magnetopause moved inward).

The magnitude of ion bulk velocity was about 500 km/s. We consider the $V_\perp(H+) = 0$ frame since the isolated flux tubes were convecting in this frame. The O+ energy in the spacecraft frame was above 10 keV (Panel f1), while in the $V_\perp(H+) = 0$ frame, the O+ energy was above ~1 keV (Panel g1). Although the observed O+ population in these flux tubes was likely the magnetospheric energetic O+, the flux was mainly distributed between 8-20 keV which was a narrower energy band than that observed in the Alfvén wing filaments between 14:05-14:24 UT.

The observed flux tubes consisted of CFLs. The O+ fluxes were primarily distributed within two pitch angle ranges: 0°-60° and 120°-180° (Panel k1). This distribution can be interpreted similarly to the newly-formed and earlier-formed CFLs discussed in Section 3.2. Specifically, the O+ ions with pitch angles directed in the nearly parallel or anti-parallel directions propagated more rapidly along the field lines than those directed in the perpendicular direction, leading to a predominance of parallel/anti-parallel O+ ions on the flux tubes.

The thicknesses of the flux tubes were slightly larger than or comparable to the O+ cyclotron scales. The flow speed perpendicular to the field lines (i.e., $V_\perp(H+)$ averaged over the entire interval 15:00-15:22 UT) was about 400 km/s. The timescales of the observed nine isolated flux tubes (highlighted by the magenta arrows in Figure 4h1) were 4, 13, 3, 25, 8, 3, 3, 28, and 17 seconds specifically. By multiplying the timescales and the perpendicular flow speed, the corresponding thicknesses of these flux tubes were 1600, 5200, 1200, 10000, 3200, 1200, 1200, 11200, and 6800 km respectively. The gyroradius of a 20-keV O+ with 60° pitch angle in a 40-nT field is about 1800 km. O+ fluxes were observed over a broader region than the flux tubes because of the O+ finite gyroradius. Satellites would observe O+ ions that were gyrating around the field lines at the outermost shell of flux tube before/after they observed the flux tube. The fact that the observed O+ ions had a significant parallel or anti-parallel velocity component suggests that they were prone to loss, especially when the flux tube had a complex structure comparable to the O+ cyclotron scales. For instance, if the flux tube was sharply kinked, O+ ions that were originally executing their helical motion inside the tube might shoot beyond the tube and end up in the solar wind. It is worth noting that because of the same reason, the observed O+ in the solar wind (e.g., between 15:06-08 UT) might also come from other flux tubes which were not in the path of MMS.

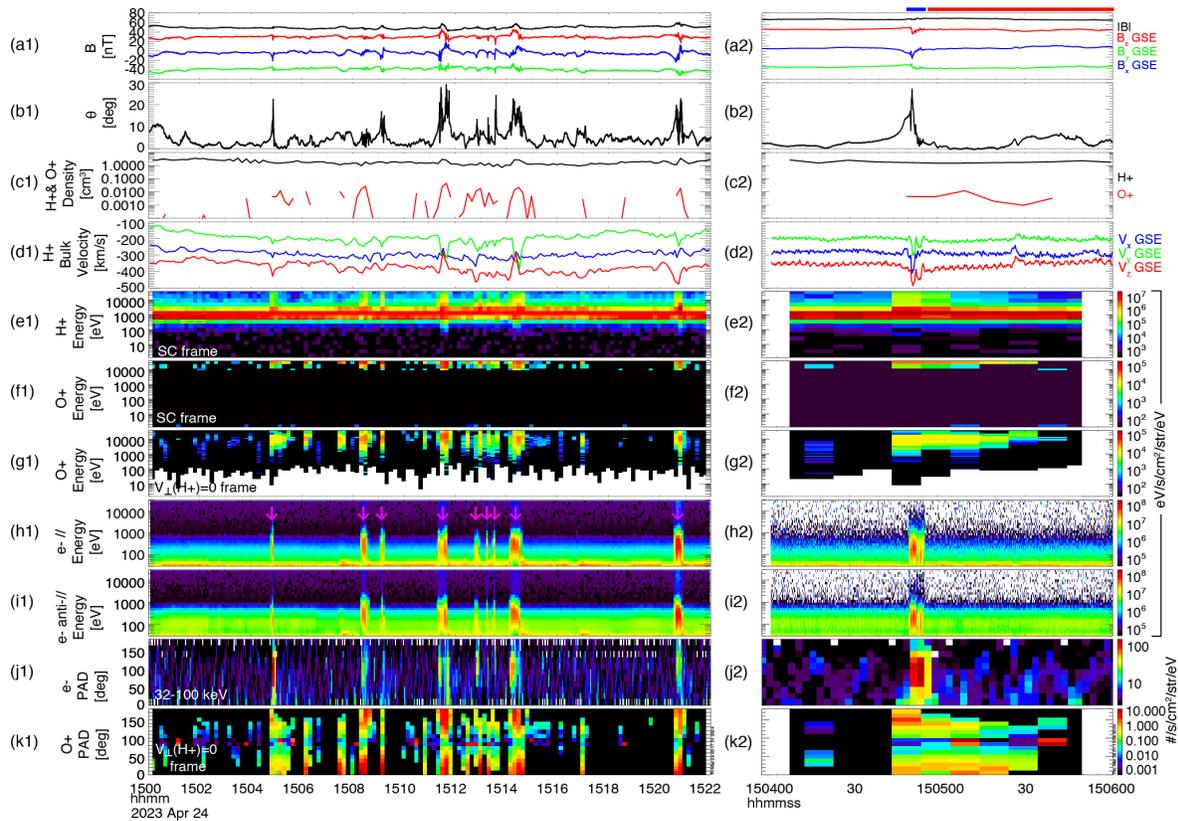

Figure 4 Observations between 15:00-15:22 UT: Left column – (a1) magnetic field, (b1) deviation angle of magnetic field from the background IMF, (c1) H+ and O+ densities, (d1) ion bulk velocity, (e1) H+ and (f1) O+ energy fluxes in the SC frame, (g1) O+ energy flux in the frame, (h1) parallel and (i1) anti-parallel electron energy fluxes to the local magnetic field, (j1) electron PAD with energy of 32-100 keV, (k1) O+ PAD in the frame. Right column – (a2)-(k2) zoom-in plots of (a1)-(k1) between 15:04-06 UT. Blue and red bars highlight the intervals containing primarily closed field lines and IMF respectively. The background IMF is obtained by averaging over 15:00-15:22 UT for (b1) and 15:05:05-15 UT for (b2).

A pitch-angle dispersion in anti-parallel O+ flux was observed starting at around 15:05 UT in Figure 4(k1). The dispersion could be seen as a gradual loss of O+ flux in 120°-180°. The exact mechanism of forming the dispersion cannot be definitively determined but the kinked flux tube scenario mentioned above offers one possible explanation for the asymmetric loss in the PAD. The Panels (a2)-(k2) shows zoom-in results from 15:04-06 UT. Intervals with signatures of CFLs (blue bar) and IMF (red bar) are determined based on electron signatures in Panels (h2), (i2) and (j2). As shown in Panel b2, the flux tube field deviated from the background IMF with a maximum 28°. O+ ions could gyrate in a helical motion oscillating between the flux tube and solar wind regions. If the deviation angle between the flux tube and IMF remains the same along the tube, the parallel and anti-parallel O+ fluxes should show the same variations when MMS cutting through the solar wind after leaving the flux tube. However, if the O+ ions, initially following helical paths between the tube and the solar wind, encounter a kink (e.g., a significant change of deviation angle) on the scale of their gyromotion, they might fail to return to the tube and be lost to the solar wind. Consequently, the parallel and antiparallel O+ fluxes could show different variations if one direction was influenced by a kink before reaching MMS. The observed gradual loss in nearly anti-parallel flux along the MMS path, compared to the relatively unaffected parallel flux, indicates that this mechanism was likely at play.

## 4. Summary and Conclusion

We study the O+ characteristics during the Alfvén-wing state of the magnetosphere induced by the CME event on Apr 24, 2023 using MMS observations. MMS was located in the dayside pre-noon sector near the southern cusp at [10.7, -7.9, -6.8] $R_E$ in the GSE coordinates. This is the first time that O+ is observed in Earth's Alfvén wing. We list summary and discussion points below:

(1) The newly-formed CFLs show less intense O+ flux with mixed parallel and anti-parallel patterns to the local magnetic field, while the CFLs that were formed earlier have intense O+ flux in mainly anti-parallel direction.

(2) The O+ ions along with energetic protons and electrons escape from the Earth into the unshocked solar wind. The fate of these escaping O+ ions remains uncertain, as they may either propagate with the MC or return to the magnetosphere.

(3) The O+ pitch-angle dispersion is observed associated with magnetic flux tubes, indicating a potential loss to the solar wind. These flux tubes appeared within the solar wind during the recovery phase of the Alfvén wing state, when the solar wind returns to super-Alfvénic speeds with the Alfvén Mach numbers between 1 and 1.5. This may indicate a subsequent loss mechanism just after the Alfvén wings disappear.

**Acknowledgments**

HL would like to express gratitude to Dr. Ying Zou for the valuable discussions. This study is supported by the NASA MMS Mission. HL acknowledges partial support from NASA grant 80NSSC24K0388. HL and HG acknowledge partial support of NSF grant AGS2247718. MMS data are publicly available at https://lasp.colorado.edu/mms/sdc/public/.